# A Fast Finite Field Multiplier for SIKE

Yeonsoo Jeon and Dongsuk Jeon

**Abstract**— Various post-quantum cryptography algorithms have been recently proposed. Supersingluar isogeny Diffie-Hellman key exchange (SIKE) is one of the most promising candidates due to its small key size. However, the SIKE scheme requires numerous finite field multiplications for its isogeny computation, and hence suffers from slow encryption and decryption process. In this paper, we propose a fast finite field multiplier design that performs multiplications in GF(p) with high throughput and low latency. The design accelerates the computation by adopting deep pipelining, and achives high hardware utilization through data interleaving. The proposed finite field multiplier demonstrates 4.48× higher throughput than prior work based on the identical fast multiplication algorithm and 1.43× higher throughput than the state-of-the-art fast finite field multiplier design aimed at SIKE.

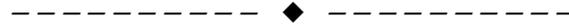

## 1 INTRODUCTION

The speed of technological advances in quantum computing is rapidly increasing. For instance, IBM has successfully factorized 4,088,459 using its 5-qubit quantum computer [1], setting a new record. It is well known that public key crptography can be easily decrypted by running Shor's algorithm on the quantum computers [2]. Therefore, post quantum cryptography (PQC) is an active research field due to its secrecy even for quantum computers.

The National Institute of Standards and Technology (NIST) started accepting PQC candidates in 2017. The result of the first round was announced in 2019 [3] and the second round is currently in progress. While there are many promising candidates such as lattice-based and code-based cryptography, isogeny-based cryptography is a strong candidate due to its advantages such as small key size and forward secrecy which is similar to conventional elliptic curve cryptography.

In 2011, the supersingluar isogeny Diffie-Hellman key exchange (SIKE) based on isogeny-based cryptography was proposed [4]. SIKE is expected to be suitable for embedded systems with limited hardware resource due to small key size. However, it requires numerous isogeny operations including modular multiplications in GF($p^2$) for encryption and decryption, resulting in slow computation. The authors in [5] showed that the secrecy of the SIKE is significantly better than what was originally estimated, which effectively speeds up the computation to meet the NIST security requirements. However, SIKE still exhibits slower encryption and decryption speed compared to other candidates.

The dominant operation in the SIKE is the multiplication of two elements in GF(p), where p is a 751-bit or smaller prime. The authors in [6] suggested the efficient finite field multiplication (EFFM) algorithm to accelerate multiplication by representing field elements in GF(p) in radix R=$2^{\alpha/2} \times 3^{\beta/2}$, reducing the cost down to 1/3. However, the expected performance improvement is limited due to long critical paths in the multiplier. The authors in [7-11] proposed a hardware accelerator for SIKE, where high-radix Montgomery multiplication is employed for faster multiplication. The Montgomery multiplication converts integers into the Montgomery domain before multiplication, replacing divisions with simple shift operations. In addition, the scheme splits the multiplication of two 2N-bit operands into a set of smaller multiplications, which enables higher operating frequency. This results in overall performance improvements despite of more clock cycles compared to EFFM.

Later, the authors in [13, 14] further improved EFFM and proposed two finite field multiplication algorithms: FFM1 and FFM2. FFM1 utilizes the property that (p - a)(p - b) is congruent with a × b (mod p) to further reduce the number of additions and precomputed values. On the other hand, FFM2 employs the property that the primes in SIKE are in the form of f×$2^{\alpha}$×$3^{\beta}$ - 1 and a × b is congruent with q+r where q and r are the quotient and remainder of a × b divided by p ± 1, respectively, in order to efficiently implement the modular operation.

These algorithms outperform EFFM in terms of throughput by 6.56× and 6.79×, respectively. However, even those variants still have longer critical paths than the Montgomery multiplication, forcing hardware to operate at a low 25MHz clock frequency. In other words, the time it takes to multiply in a finite field is still longer than that of the high-radix Montgomery multiplication.

The authors in [15] further improved FFM1 algorithm. They introduced an improved Barrett Reduction (IBR) and utilized it in hardware implementation. Also, they presented unconventional radix method to perform a modular multiplication. As a result, they could reduce computational resources required to perform a modular multiplication and also reduced the execution time. However, the method does not support arbitrary SIKE primes. That is, the method can only be used for the primes that have the form of f×$2^{2\alpha}$×$3^{2\beta}$ − 1, where the exponents are limited to even numbers.

Recently, the authors in [16] presented a high performance modular multiplication (HFFM) algorithm. The algorithm is similar to FFM1, but it employs a different radix system and overcomes the limitation that α and β must be even numbers. Moreover, it adopts interleaving method, resulting in higher throughput than prior works.

---

• *Y. Jeon and D. Jeon are with the Seoul National University, Seoul, South Korea 08826.*

In this paper, we propose a new hardware architecture for fast finite field multiplications aimed at SIKE using the FFM2 algorithm. We deeply pipeline the multiplier and thus reduce the critical path delay, which increases the maximum clock frequency to 100MHz. In addition, we employ data interleaving in the multiplier to maximize hardware utilization, similar to [16]. By processing two independent input pairs using a single pipelined multiplier, the latency is nearly halved and achieves 4.48× throughput than the prior FFM2 implementation in [14].

Our paper is organized as follows. In section 2, we briefly introduce the background of modular reduction methods. In section 3, we propose a new finite field multiplier architecture. In section 4, we present the implementation results and compare the performance with prior works. Finally, section 5 concludes the paper.

## 2 BACKGROUND

### 2.1 Primes in SIKE

SIKE scheme uses special primes in the form of $f \times 2^\alpha \times 3^\beta - 1$, where the primes are typically very big numbers [5, 17]. The proposal [17] suggests primes such as SIKEp434, SIKEp503, SIKEp610, and SIKEp751. The number in the name represents the number of bits of the prime, where SIKEp751 is the biggest prime targeting NIST security level 5. To ensure the secrecy of the scheme, SIKE carefully chooses the parameters α and β such that $2^\alpha \approx 3^\beta$. In this paper, we target the largest prime (SIKEp751) that provides the highest securacy.

The crucial part in SIKE is calculating $a \times b$ (mod p) where $a$ and $b$ are typically very large numbers. Below are existing algorithms that can efficiently calculate modular multiplication.

### 2.2 High-Radix Montgomery Multiplication

Consider multiplying two finite field elements $a$ and $b$ in GF($p$) so that $c = a \times b$. Since the field is closed under multiplication, the remainder of $a \times b$ divided by $p$ represents the element c in the field. Therefore, typically $a$ is multiplied by $b$ in a trivial way and the remainder calculation follows. However, the division is very expensive operation in hardware.

Montgomery reduction algorithm reduces the hardware cost by converting the operands into the Montgomery domain before multiplication. To be specific, with some special integer $R$, usually power of 2 in computers, it converts $a$ and $b$ to $aR$ and $bR$, which are Montgomery representations of $a$ and $b$, respectively. Then, instead of multiplying $a$ and $b$, the algorithm multiplies $aR$ and $bR$ and divides the result by $R$ to get $abR$, which is also a Montgomery representation of $c = ab$, the result of ordinary multiplication.

The original algorithm was first proposed in [18], and an improved high-radix Montgomery multiplication algorithm (Algorithm 1) that performs multiplication and modulo operation recursively was presented in [19].

The authors in [20] implemented the high-radix Montgomery multiplication in hardware using a systolic architecure. The calculation is splited into smaller multipliers, significantly reducing the critical path delay. Thus, although it takes more clock cycles to compute finite field multiplication than Montgomery reduction, the clock frequency is largely increased, resulting in shorter total computation time.

---

**Algorithm 1**: High-Radix Montgomery Multiplication [19]

Input:
$A = \sum_{i=0}^{m+2} (2^k)^i a_i, a_i \in \{0, 1, ..., 2^k - 1\}, a_{m+2} = 0$
$B = \sum_{i=0}^{m+1} (2^k)^i b_i, b_i \in \{0, 1, ..., 2^k - 1\}$
$MM' = -1 \ (mod \ 2^k)$
$\bar{M} = (M' mod \ 2^k)M$
$\bar{M} = \sum_{i=0}^{m} (2^k)^i \bar{m}_i, \bar{m}_i \in \{0, 1, ..., 2^k - 1\}$
$A, B < 2\bar{M}$
$4M < 2^{km}$

Output: $S_{m+3} = ABR^{-1}(mod \ M)$

1. $S_0 = 0$;
2. for $i$ in 0 to $m+2$ loop
3. $q_i = S_i \ mod \ 2^k$;
4. $S_{i+1} = (S_i + q_i \bar{M})/2^k + a_i B$;
5. endloop
6. **return** $S_{m+3} = ABR^{-1} \ (mod \ M)$;

---

### 2.3 Barrett Reduction

Barrett reduction [21] is an algorithm that takes advantage of precomputation. Barrett reduction can perform reductions fast if the modular is fixed. If we know the quotient $q$ of $a$ divided by $p$ exactly, we can get the remainder by subtracting $q \times p$ from the original number $a$. However, one needs to find the quotient $q$ as accurately as possible to avoid error. The quotient $q$ can be directly calculated by dividing $a$ by $p$. However, since division typically costs more hardware resource than multiplication, division is replaced by multiplication with its inverse. If $1/p$ is accurate enough, then one can obtain the $q$ accurately.

The Barrett reduction is detiled in Algorithm 2. For well-chosen k that satisfies $2^{k-1} > p$, we take $1/p = x/2^k$ as an approximation. In order for $x$ to be an integer, usually $x = floor(2^k/p)$ is chosen. Since the approximated value of $x/2^k$ is less than or equal to $1/p$, the error of the approximated $q$ is $e = 1/p - x/2^k$. To obtain a correct result, $ae$ must be less than 1. Thus, $a < 2^k$ is a sufficient condition and this can be entailed by choosing a proper $k$.

---

**Algorithm 2**: Barrett Reduction [21]

**Input:** $a, p$, parameter $k, x = \lfloor 2^k/p \rfloor$
**Output:** $a \ (mod \ p)$

1. $q = a \times x \gg k$;
2. $r = a - q \times p$;
3. **if** $r \geq p$ **then**
4. $r = r - p$;
5. $q = q + 1$;
6. **end**
7. **return** $q, r$;

## 2.4 FFM2 Algorithm

The authors in [14] further optimized the Barrett reduction for the primes used in SIKE and proposed FFM2 algorithm. The SIKE prime $p$ is expressed as $p = T - 1$, where $T = 2^a \times 3^\beta$. For $C = A \times B$, consider the quotient $q'$ and the remainder $r'$ of $c$ divided by $T = p + 1$; then, we obtain the following equation:

$$C = q'T + r' = q'p + q' + r' \equiv q' + r' \pmod{p} \quad (1)$$

However, finding $q'$ and $r'$ is the most difficult step in the algorithm because of the division. Since $T = 2^a \times 3^\beta$, the algorithm can first perform division by $2^a$, leaving smaller division by $3^\beta$. Division by $3^\beta$ can be realized using Barrett reduction, and this results in a multiplication with a narrower bitwidth than the original method. Equation 2 shows how the separated divisions are combined. The final results are obtained from $q' = q_2$, and $r' = (r_2 \times 2^a) + r_1$.

$$C = q_1 \times 2^a + r_1 = 2^a(3^\beta \times q_2 + r_2) + r_1 \quad (2)$$

---
**Algorithm 3**: FFM2 Algorithm [14]
  **Input:** $A, B \in F_p$, $p = f \cdot 2^a 3^b \pm 1$
  **Output:** $C = A \times B \pmod{p}$
1. $C = A \times B$;
2. $q_1 = C / 2^a$;
3. $r_1 = C \% 2^a$;
4. $q_2, r_2 = \text{Barrett}(q_1, f \cdot 3^b)$;
5. $r_2 = r_2 \ll 2^a + r_1$;
6. **if** $p == f \cdot 2^a 3^b - 1$ **then**
7.   $C = q_2 + r_2$;
8.   if $C > p$ then
9.     $C = C - p$;
10.   end
11. end
12. else
13.   $C = r_2 - q_2$;
14.   If $C < 0$ then
15.     $C = C + p$;
16.   end
17. end
18. return $C$;
---

## 2.5 Number of Multiplications in Various Algorithms

The complexity of various algorithms is analyzed in [22], where the cost of finite field multiplication is calculated as the number of $k$-bit multiplication instructions. Let $k$ be the number of bits that is multiplied at a time and let $n$ be the number of bits in modulus $p$. Then, $p$ can be split into $m$ $k$-bit sub-blocks where $n \leq m \times k$.

The high-radix Montgomery multiplication algorithm costs $m^2 + m$ multiplication instructions. The authors in [12] notice that SIKE primes have a special form and further reduce the number of operations to nearly half. Specifically, for SIKE prime, the precomputed value $M'$ in Algorithm 1 is equal to 1. Therefore, the optimized high-radix Montgomery multiplication algorithm requires $m^2/2$ multiplication instructions.

The multiplication cost of Barrett reduction is estimated as $m^2+4m+1$. In FFM2 algorithm, the input is divided by $2^a$ before applying Barrett reduction, reducing the cost to

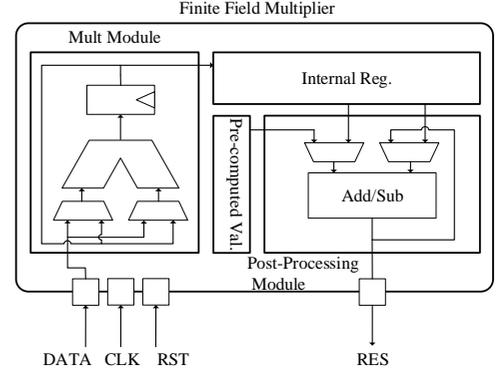

Figure 1. The overall architecture of the proposed fast finite field multiplier.

$5/8m^2 + 13/4m + 1$.

In all cases, it is advantageous to choose a small $m$, which is equivalent to choosing a large $k$, to reduce the multiplication cost. If we use the same $m$ for all algorithms, high-radix Montgomery multiplication exhibits the lowest cost, as concluded in [22]. However, the optimal value of $m$ may differ for each algorithm. Since application-specific integrated circuits (ASICs) can be designed for multiplication with arbitrary widths, the execution time of a multiplication instruction depends on the value of $m$. Smaller $m$ reduces the number of multiplications, but it simultaneously increases the delay of critical paths in the multiplier. Therefore, there exists an optimal value for $m$ that minimizes the total computation time. For fair comparisons between ASIC implementations of algorithms, the performance must be measured as the total computation time instead of instruction counts.

## 3 FAST FINITE FIELD MULTIPLIER

The authors in [14] demonstrated a finite field muliplier design based on the FFM2 algorithm on an FPGA platform, which is used as a baseline in this paper. The finite field multiplier is composed of a multiplier, an adder, and a subtractor. The authors presented two designs using different $k$ values (385 and 193), where the design with larger $k$ demonstrated shorter computation time. However, as $k$ is still very large in those designs, long critical paths in the multiplier severely limit the clock frequency. In this chaper, we propose a fast finite field multiplier design, which significantly improves the throughput through deep pipelining and data interleaving.

### 3.1 Proposed Hardware Architecture

As the large multiplier is the performance bottleneck in the baseline design [14], we propose to employ a deeply pipelined multiplier so as to raise the calculation performance. Pipelining is a commonly used design technique to accelerate calculation by shortening the critical path delay [23]. However, deep pipelining may incur hardware underutilization. Hence, we also propose a data interleaving technique to boost the utilization of the multiplier, which will be detailed in the next section.

Fig. 1 describes the overall architecture of the proposed fast finite field multiplier. The design supports $2N \times 2N$-bit finite field multiplication and is composed of a reconfigurable multiplier, a ROM storing precomputed values, a post-processing module, and a controller.

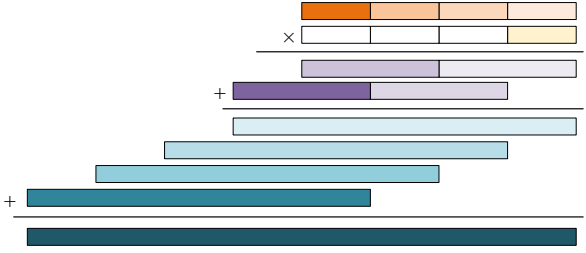

Figure 2. The datapath of the $N$-bit multiplier.

The multiplication process is as follows: the design takes two $2N$-bit values $A$ and $B$, and first performs a $2N \times 2N$-bit multiplication using the $N \times N$-bit multiplier, which corresponds to line 1 in Algorithm 3. The results is splitted into the quotient $q_1$ and remainder $r_1$ of division by $2^\alpha$ (line 2 and 3). $q_1$ is then divided by $3^\beta$ again using Barrett reduction, resulting in the quotient $q_2$ and remainder $r_2$. The Barrett reduction requires a $2N \times 3N$-bit multiplication and a $2N \times N$-bit multiplication, which are again realized by the $N \times N$-bit multiplier. The post-processing module is responsible for the rest of the algorithm (line 5 through 18 in Algorithm 3). The ROM stores the pre-computed values for $2^\alpha$, $3^\beta$, $p$, and Barrett reduction. The controller is implemented as a finite state machine (FSM) and controls the datapath so that the $N \times N$-bit multiplier and other modules are assigned for each algorithm step appropriately.

### 3.2 Reconfigurable Multiplier

The reconfigurable multiplier adopted in the design is essentially an $N \times N$-bit multiplier with additional reconfigurable accumulation paths to realize multiplications with different bitwidths. Since a vanilla $N \times N$-bit multilier would have a very long critical path, we divide the multiplier into 5 stages by decomposing the multiplier into 16 $N/4 \times N/4$-bit multipliers. Those small multipliers generate partial products from $N/4$ bits of each operand in the first stage, as shown in Fig. 2, and additional adders accumulate those partial products to obtain the final $2N$-bit result in the remaining 4 pipeline stages. However, even the small $N/4 \times N/4$-bit multipliers still have a long critical path delay and hence we pipeline those multipliers into 3 stages to further enhance the performance, pipelining the reconfigurable multiplier into 9 stages including one stage for buffering.

On the other hand, pipelining also increases the latency in terms of the number of clock cycles. For instance, 3-stage pipelined $N/4 \times N/4$-bit multiplier would take two more clock cycles for multiplication compared to a non-pipelined version. However, since pipelining reduces the critical path delay and lets the system operate with higher clock frequency, the latency increase in terms of absolute time is largely suppressed. In addition, the algorithm can process multiple N-bit multiplications in parallel, the latency is further reduced. For instance, if a $2N \times 2N$-bit multiplication is decomposed into 4 independent $N \times N$-bit multiplications, they can be serially processed by a 9-stage pipelined $N \times N$-bit multiplier, which takes 9+3=12 cycles to finish the multiplication. The latency increase would only be 8 cycles compare to a single-stage $N \times N$-bit multiplier that takes 4 cycles to process all 4 $N \times N$-bit multiplications. However, since the clock frequency is increased by 9× through pipelining (excluding pipelining overheads), the latency in absolute time is actually decreased by 3×, which is confirmed by experimental results demonstrated in the next section.

There are three different multiplications in the FFM2 algorithm: $2N \times 2N$-bit, $3N \times 2N$-bit, and $2N \times N$-bit multiplications. Therefore, a single $N \times N$-bit multiplier can accommodate all three multiplications by decomposing them into $N \times N$-bit multiplications. It would take 4, 6, and 2 clock cycles to accomplish those multiplications, respectively.

Note that these multiplications cannot be processed in parallel and must be executed sequentially. For example, $3N \times 2N$ multiplication for Barrett reduction (line 4 in Algorithm 3) can only be started after the $2N \times 2N$ multiplication is done (line 1) and followed by division (line 2 and 3). Fig. 3 describes how the reconfigurable multiplier processes those multiplications in detail.

### 3.3 Data Interleaving

While the deeply pipelined multiplier largely increases the throughput, the hardware can be underutilized. For instance, as explained in the previous section, the $3N \times 2N$-bit multiplication can start only after the $2N \times 2N$-bit multiplication is completed, resulting in bubbles in the pipeline stages. For instance, while processing $2N \times 2N$-bit multiplication, the utilization drops to 33.3% (4 $N \times N$-bit multiplications processed in 9+3=12 cycles). Similarly, the utilization becomes 42.9% and 20% for $3N \times 2N$-bit and $2N \times N$-bit multiplications, respectively. Therefore, we propose a data interleaving method to maximize the hardware utilization. Specifically, we modify the design to process two independent sets of data in parallel and nearly double the hardware utilization, making the utilization back to 83.8% for the $3N \times 2N$-bit multiplication.

Fig. 4 shows an example of $2N \times 2N$-bit multiplication when the data interleaving is applied along with the timing diagram. Suppose that two sets of data are multiplied, $(a_1, b_1)$ and $(a_2, b_2)$. The $N \times N$-bit multiplier takes two inputs from the first data set for 4 clock cycles, and the next set is fed into the multiplier for the following 4 clock cycles. The results corresponding to the first set appear at the output port starting at clock cycle 9. For example, the first output corresponds to $a_{11} \times b_{11}$, the second output is equal to $a_{11} \times b_{12}$, and so on. Other multiplications such as $3N \times 2N$-bit and $N \times 2N$-bit are also interleaved in a similar manner. As a result, the interleaving method makes the reconfigurable multiplier finish all three multiplications for two independent sets of integers within 50 clock cycles. The results for the first and second sets are generated at clock cycle 43 and 50, respectively.

Being able to process two independent data in parallel is beneficial to SIKE. Like other cryptography systems, the SIKE scheme is also defined over $GF(p^2)$. If the hardware can compute two data simultaneously, it can accelerate multiplication over $GF(p^2)$ by processing multiplications of real and imaginary parts in parallel.

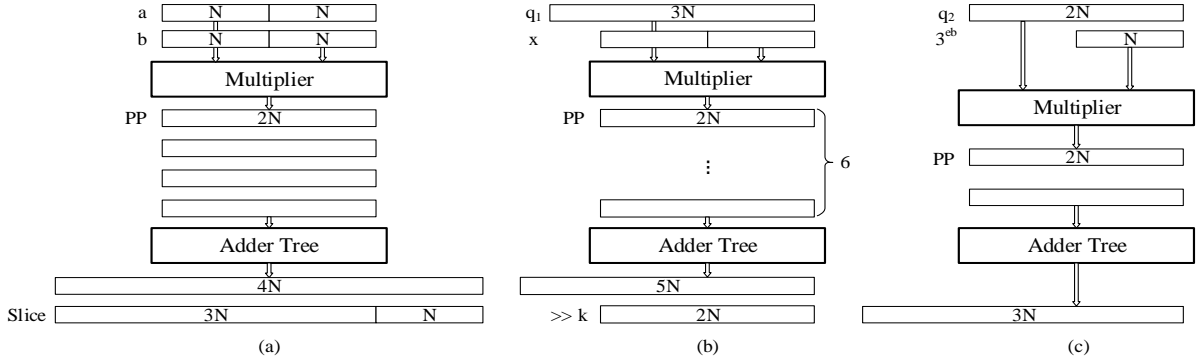

Fig. 3. Reconfigurable multiplier processing (a) $2N \times 2N$, (b) $3N \times 2N$, and (c) $2N \times N$-bit multiplications.

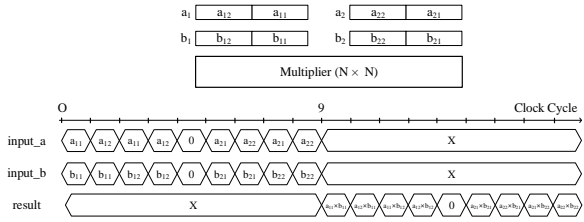

Fig. 4. Timing diagram of $2N \times 2N$-bit multiplication.

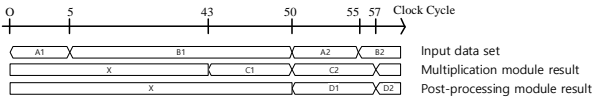

Figure 5. Timing diagram of the entire system.

Finally, the output of the reconfigurable multiplier is sent to the post-processing module whenever the multiplier completes its operation. The post-processing module will be described in the next section.

### 3.4 Post-Processing Module

The post-processing module processes the interim result from the reconfigurable multiplier and generates the final product, corresponding to line 5 through 18 in Algorithm 3. While the post-processing module processes the interim results, the reconfigurable multiplier starts working on a new set of integers, maximizing hardware utilization.

The post-processing module recombines the intermediate results from the reconfigurable multiplier to produce the final result. It takes a total of 7 clock cycles to complete the calculation. Fig. 5 shows the timing diagram of the entire system. If the first input arrives at T = 0 and the second input arrives at T = 5, the reconfigurable multiplier returns the first and second sets of output at T = 43 and 50, respectively. After the post-processing module processes the first intermediate result, the final product of the first set is produced at clock cycle 50, at which the second intermediate result becomes ready. Hence, the post-processing of the second intermediate result is performed immediately, delivering the final product of the second set at clock cycle 57. At cycle 50, the reconfigurable multiplier finishes processing the second set of data. Therefore, it moves to the next batch of data, which allows the system to process two sets of data every 50 clock cycles (i.e., throughput is 25 cycles per multiplication), whereas the latency is 57 cycles. Note that the post-processing module is only activated twice every 57 cycles and its utilization is relatively low. However, since the system is dominated by the reconfigurable multiplier in terms of both power and area, the overhead is negligible.

## 4 EXPERIMENTAL RESULTS

To evaluate the proposed finite field multiplier, we implemented the design using Xilinx Vivado 2019.1. We compare the proposed multiplier design against the baseline in [14] which adopts the identical FFM2 algorithm and other prior works [15, 16] based on SIKEp771. For fair comparisons with [14], we set the target device to Xilinx Kintex-7 xc7k325tffg900-2 which is the device used for experiments in [14]. We also compare the proposed design to the finite field multiplier design in [11], which targets SIKEp751 and proposed full SIKE scheme in FPGA. We set the target device to Virtex-7 xc7vx690tffg1157-3 board and used the same prime. The comparison results are shown in Table I.

Compared to the baseline [14], the proposed multiplier was successfully synthesized and implemented with significantly higher clock frequency of 100MHz, which directly translates to largely improved throughput, at the expense of 2.71× more FFs due to deep pipelining. The amout of other resources such as LUT and DSP is nearly identical as expected.

Since our design utilizes a 9-stage pipelined multiplier, more clock cycles are required to complete a multiplication. However, as discussed in Section 3.2, higher clock frequency and parallel processing reduce the latency in absolute time by 1.96×. In addition, data interleaving technique enables the system to process two sets of data at a time, making our design achieve 4.48× higher throughput than the baseline.

The design in [15] demonstrates higher throughput and lower latency, but it occupies large area and relies on the algorithm that only supports a limited set of primes (i.e., the exponent must be even). Hence, the hardware is not versatile. On the other hand, the proposed design can be used for any SIKE primes.

TABLE I
COMPARISONS WITH PRIOR WORKS

| Algorithms | [14] | [15] | [16] | This work | [11] | Thiswork |
|---|---|---|---|---|---|---|
| Target Prime | SIKEp771 | | | | SIKEp751 | |
| Platform | Kintex-7 | Vertex-7 | Vertex-7 | Kintex-7 | Vertex-7 | Vertex-7 |
| FFs | 11632 | 38976 | 10680 | 31541 | N/A | 31476 |
| LUTs | 33051 | 63173 | 11007 | 31014 | N/A | 30970 |
| DSPs | 529 | 729 | 144 | 576 | 128 | 576 |
| Frequenc (MHz) | 25 | 60 | 152 | 100 | 167 | 111 |
| Latency (cycles) | 28 | 18 | 66 | 57 | 100 | 57 |
| Latency (ns) | 1120 | 306 | 410 | 570 | 597 | 513 |
| Throughput (cycles/mult) | 28 | 1 | 54 | 25 | 69 | 25 |
| Throughput (MIPS) | 0.89 | 58.8 | 2.8 | 4 | 2.42 | 4.44 |

Compared to the design in [16], our design shows 1.86× longer latency, but achieves 1.43× higher throughput while consuming more hardware resources. This limitation is resulted from the property of the algorithm; FFM1 and HFFM require less multiplication operations than our baseline algorithm, FFM2.

The authors in [11] choose the high-radix Montgomery multiplication algorithm to multiply two numbers. The algorithm is fundamentally different from other algorithms used in [14, 15, 16]. The Montgomery algorithm requires more clock cycles to finish a finite field multiplication compared to the FFM2 algorithm, but it relies on narrow bitdwidth multiplications, resulting in shorter critical path delay and overall computation time. Compared to the design in [11], the proposed design requires more DSPs, resulting in larger area. This is because our design requires 95×95-bit multiplications. The DSP48E1 slice in the target FPGA device can afford upto 25×18-bit two's complement multiplication, whereas our design requires 95 × 95-bit multiplications. The synthesis tool automatically groups multiple DSP blocks to implement a large multiplier. For example, for a symmetric unsigned multiplication, one DSP block can handle 17×17-bit multiplication. Grouping two blocks enables 24×24-bit multiplication and grouping four blocks raises the capacity to 34×34-bit. As a result, 35 DSP blocks are required to accommodate 95×95-bit multiplications adopted in the proposed design. Since our design employs 16 multipliers, it uses 560 DSP blocks to implement mulipliers. On the other hand, the design in [11] uses 128 DSP blocks in total for a finite field multiplier. This is because the authors in [11] decompose the operation into multiplications with narrower bitwidth ($k = 24$). Thus, it only needs 2 DSP blocks for each multiplier.

Nevertheless, the proposed architecture exhibits improvement in latency and throughput compared to [11]. While the prior work takes 100 clock cycles to generate the first result (which is equivalent to the latency), our design only takes 57 clock cycles to generate the outcome. Moreover, due to data interleaving technique, our architecture also demonstrates throughput improvement. While the prior design takes 69 clock cycles per multiplication, the proposed design can compute 2 sets of integers in 50 clock cycles, making the throughput 25 clock cycles per multiplication.

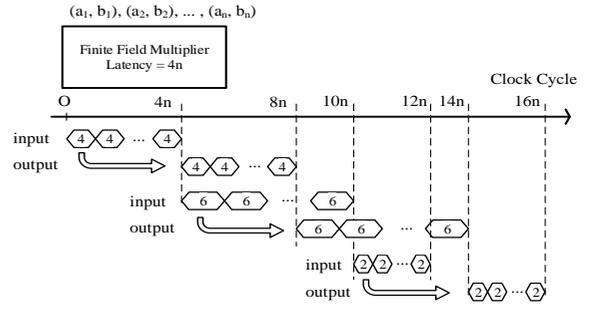

Fig. 6. Timing diagram for the scenario that the multiplier is fully utilized.

Furthermore, our design operates at 111MHz clock frequency, resulting in the calculation time of 225 ns for each multiplication and 4.44 MIPS throughput through deep pipelining and data interleaving. Therefore, the result reveals that our architecture improves the performance by 1.83× and reduces the latency by 1.16×.

In summary, our design can multiply arbitrary SIKE primes, and has higher throughput compared to other accelerators. Our design performs achieves higher throughput than that of the design in [14], which adopts the identical algorithm. Also, our design has higher throughput compared to other designs in [11, 15, 16].

Based on these experimental results, the proposed hardware has a potential to speed up the existing FPGA accelerators for SIKE. The authors in [11] show that addition over GF(p) requires 2 clock cycles and interleaved multiplication requires 69 clock cycles to complete its operation, confirming that the multiplication is the computational bottleneck. In addition, the authors in [10] counted the number of operations for a SIDH key exchange, and showed that the number of multiplications is twice the number of additions. Therefore, one can conclude that the finite field multiplication dominates the performance of the SIDH scheme. Given that the proposed finite field multiplier supports fast multiplication, the key exchange process will be significantly accelerated as well.

In addition, the proposed finite field multiplier can be further optimized to achieve full utilization of hardware,

while the current design has 83.8% multiplier utilization as described in Section 3.2. Suppose we are multiplying $n$ independent set of integers $(a_1, b_1)$ through $(a_n, b_n)$ simultaneously (i.e., interleaving $n$ data sets). If the finite field multiplier has $4n$ latency, the design can fully utilize the hardware multiplier as shown in the timing diagram in Fig. 6. First, for the first multiplication in the FFM2 algorithm ($2N \times 2N$-bit), the $n$ sets of input data are transferred to the multiplier over $4n$ cycles since the multiplier takes 4 cycles to process each set. After $4n$ cycles, which is the latency of multiplier, the multiplication result for the first input set becomes ready. Then, the next multiplication in the FFM2 algorithm ($3N \times 2N$-bit) begins immediately. Since the second multiplication takes 6 cycles for each set, the hardware processes this step for $6n$ cycles and the multiplier becomes available at t = $10n$. Since the first output of the second multiplication is generated at t = $8n$, these outputs are buffered until t = $10n$. Finally, for the next $2n$ cycles, the last multiplication ($N \times 2N$-bit) in the algorithm is performed. The final results are produced from t = $14n$ to t = $16n$. With this scheme, the multiplier is fully utilized and the throughput of the system is improved to $14n/n$ = 14 cycles per multiplication.

The throughput of the finite field multiplier is determined by the pipelined stages of small multipliers and the number of those small multipliers in the finite field multiplier. Therefore, if we choose the bitwidth of the operands and the pipelining stage appropriately, the system can be configured optimally and achieve the maximum throughput of 14 cycles per finite field multiplication. For example, only with 100 MHz clock frequency, the system can reach up to 7.14 MIPS.

## 5 CONCLUSIONS

In this paper, we propose a fast finite field multiplier that can be used in isogeny-based cryptography such as SIKE. Through deep pipelining and data data interleaving techniques, the proposed design demonstrates superior performance compared to prior works. Specifically, deep pipelining addresses long critical path delays in conventional FFM2 implementations and enhances overall throughput by allowing significantly higher operating frequency. In addition, interleaving input data resolves low hardware utilization due to deep pipelining as well as further improves the throughput.

Experimental results confirm that the proposd design achieves 1.96× lower latency and 4.48× higher throughput than prior work implementing the identical FFM2 algorithm. In addition, our design exhibits 1.43× higher throughput compared to state-of-the-art design based on HFFM algorithm.

## REFERENCES

[1] A. Dash, D. Sarmah, B. K. Behera, P. K. Panigrahi, "Exact search algorithm to factorize large biprimes and a triprime on IBM quantum computer," 2018. [Online]. Available: https://arxiv.org/abs/1805.10478

[2] P. W. Shor, "Algorithms for quantum computation: Discrete logarithms and factoring," *in Proc. Symp. Found. Comput. Sci., 1994*, pp. 124–134.

[3] https://nvlpubs.nist.gov/nistpubs/ir/2019/NIST.IR.8240.pdf

[4] D. Jao and L. De Feo, "Towards quantum-resistant cryptosystems from supersingular elliptic curve isogenies," *in Proc. Post-Quantum Cryptography, 2011*, pp. 19–34.

[5] C. Costello, P. Longa, M. Naehrig, J. Renes, and F. Virdia, "Improved classical cryptanalysis of the computational supersingular isogeny problem," Cryptology ePrint Archive, Report 2019/298, https://eprint.iacr.org/2019/298.

[6] A. Karmakar, S. S. Roy, F. Vercauteren, and I. Verbauwhede, "Efficient Finite Field Multiplication for Isogeny Based Postquantum Cryptography," *Proc. Tenth International Workshop on Arithmetic of Finite Fields*, pp. 193-207, 2016.

[7] B. Koziel, R. Azarderakhsh, M. Mozaffari Kermani, and D. Jao, "Post-quantum cryptography on FPGA based on isogenies on elliptic curves," *IEEE Trans. Circuits Syst. I: Reg. Papers*, vol. 64, no. 1, pp. 86–99, Jan. 2017.

[8] B. Koziel, R. Azarderakhsh, and M. Mozaffari-Kermani, "Fast Hardware Architectures for Supersingular Isogeny Diffie-Hellman Key Exchange on FPGA," *in Progress in Cryptology - INDOCRYPT 2016: 17th International Conference on Cryptology in India*, 2016, pp. 191-206.

[9] B. Koziel, A. Jalali, R. Azarderakhsh, D. Jao, and M. Mozaffari-Kermani, "NEON-SIDH: Effcient Implementation of Supersingular Isogeny Diffie-Hellman Key Exchange Protocol on ARM," *in Cryptology and Network Security: 15th International Conference, CANS 2016*, 2016, pp. 88-103.

[10] B. Koziel, R. Azarderakhsh, and M. Mozaffari-Kermani, "A High-Performance and Scalable Hardware Architecture for Isogeny-Based Cryptography," *IEEE Transactions on Computers*, vol. 67, no. 11, pp. 1594-1609, Nov 2018.

[11] B. Koziel, A. Ackie, R. E. Khatib, R. Azarderakhsh and M. Mozaffari-Kermani, "SIKE'd Up: Fast and Secure Hardware Architectures for Supersingular Isogeny Key Encapsulation," Cryptology ePrint Archive, Report 2019/711, https://eprint.iacr.org/2019/711.

[12] C. Costello, P. Longa, and M. Naehrig, "Efficient algorithms for supersingular isogeny Diffie-Hellman," *in Proc. Annu. Int. Cryptology Conf. Advances Cryptology, 2016*, pp. 572–601.

[13] C. Liu, J. Ni, W. Liu, Z. Liu and M. O'Neill, "Design and Optimization of Modular Multiplication for SIDH," *2018 IEEE International Symposium on Circuits and Systems (ISCAS)*, May. 2018, doi:10.1109/ISCAS.2018.8351082.

[14] W. Liu, J. Liu, C. Liu and M. O'Neill, "Optimized Modular Multipliacation for Supersingular Isogeny Diffie-Hellman," *IEEE Transactions on Computers*, vol. 68, no. 8, pp. 1249-1255.

[15] J. Tian, J. Lin and Z. Wang, "Ultra-Fast Modular Multiplication Implementation for Isogeny-Based Post-Quantum Cryptography," 2019 IEEE International Workshop on Signal Processing Systems (SiPS), Nanjing, China, 2019, pp. 97-102.

[16] W. Liu, Z. Ni, J. Ni, C. Rafferty and M. O'Neill, "High Performance Modular Multiplication for SIDH," in IEEE Transactions on Computer-Aided Design of Integrated Circuits and Systems, Dec. 2019, doi: 10.1109/TCAD.2019.2960330.

[17] D. Jao, R. Azarderakhsh, M. Campagna, C. Costello, L. De Feo, B. Hess, A. Jalali, B. Koziel, B. LaMacchia, P. Longa, M. Naehrig, J. Renes, V. Soukharev, and D. Urbanik, "Supersingular Isogeny Key Encapsulation," Submission to the NIST Post-Quantum Standardization Project, 2017. [Online]. Available: https://sike.org/

[18] P. L. Montgomery, "Speeding the Pollard and elliptic curve methods of factorization," *J. Math. Comput.*, vol. 48, pp. 243–264, 1987.

[19] H. Orup, "Simplifying Quotient Determination in High-Radix Modular Multiplication," *Proc. 12th Symp. Computer Arithmetic*, pp. 193-199, 1995.

[20] T. Blum and C. Paar, "High-radix montgomery modular exponentiation on reconfigurable hardware," *IEEE Transactions on Computers*, vol. 50, no. 7, pp. 759–764, Jul. 2001.

[21] P. Barrett, "Implementing the Rivest Shamir and Adleman Public Key Encryption Algorithm on A Standard Digital Signal Processor," *Proc. Seventh Annual International Cryptology Conference on Advances in Cryptology (CRYPTO)*, LNCS vol. 263, pp. 311-323, 1987.

[22] J. Bos and S. Friedberger, "Arithmetic Considerations for Isogeny Based Cryptography," *IEEE Transactions on Computers*, vol. PP, no.99, pp. 1-99, 2018.

[23] G. Athanasiou, G. Makkas and G. Theodoridis, "High throughput pipelined FPGA implementation of the new SHA-3 cryptographic hash algorithm", *6th IEEE Int. Symp. on Communications, Control and Signal Processing (ISCCSP)*, Athens, Greece, pp. 538-541, 21-23 May 2014.